\documentclass[conference]{IEEEtran}
\IEEEoverridecommandlockouts

\usepackage{cite}
\usepackage{amsmath,amssymb,amsfonts}
\usepackage{algorithmic}
\usepackage{graphicx}
\usepackage{textcomp}
\usepackage{xcolor}
\usepackage{multirow}
\usepackage{threeparttable}
\usepackage{booktabs}
\usepackage{epstopdf}
\usepackage[T1]{fontenc}
\usepackage{aecompl}
\usepackage{cleveref}
\usepackage{tabularx}
\usepackage{nicematrix,tikz}
\usepackage{color, soul}

\def\BibTeX{{\rm B\kern-.05em{\sc i\kern-.025em b}\kern-.08em
    T\kern-.1667em\lower.7ex\hbox{E}\kern-.125emX}}

\usepackage[all]{background}
\usepackage{url}
\usepackage{lipsum}

\SetBgContents{Journal of Energy Storage https://doi.org/10.1016/j.est.2024.113502}
\SetBgPosition{current page.west}
\SetBgVshift{-0.5cm}
\SetBgOpacity{3}
\SetBgAngle{90.0}
\SetBgScale{1.8}

\usepackage{fancyhdr}
\usepackage{lipsum}

\begin{document}

\title{Transfer Learning-based State of Health Estimation for Lithium-ion Battery with Cycle Synchronization\\
}

\author{Kate Qi~Zhou,~\IEEEmembership{}
	Yan~Qin,~\IEEEmembership{Member,~IEEE,}
	Chau~Yuen,~\IEEEmembership{Fellow,~IEEE}
	\thanks{This work was supported by EMA-EP011-SLEP-001. (Corresponding author: Yan Qin)}
	\thanks{K. Q. Zhou, is with the Engineering Product Development Pillar, The Singapore University of Technology and Design. (e-mail: qi$\_$zhou@mymail.sutd.edu.sg, yan.qin@cqu.edu.cn, yuenchau@sutd.edu.sg)}
	\thanks{Y. Qin is with the Engineering Product Development Pillar, The Singapore University of Technology and Design, and Design, and School of Automation, Chongqing University,  P. R.China (e-mail: yan.qin@cqu.edu.cn)}
	\thanks{C. Yuen is with the School of Electrical and Electronics Engineering,  Nanyang Technological University (e-mail: chau.yuen@ntu.edu.sg)}}


\onecolumn
\maketitle
\begin{abstract}
Data-driven methods have gained extensive attention in estimating the state of health (SOH) of lithium-ion batteries.  Accurate SOH estimation requires degradation-relevant features and alignment of statistical distributions between training and testing datasets. However, current research often overlooks these needs and relies on arbitrary voltage segment selection. To address these challenges, this paper introduces an innovative approach leveraging spatio-temporal degradation dynamics via graph convolutional networks (GCNs). Our method systematically selects discharge voltage segments using the Matrix Profile anomaly detection algorithm, eliminating the need for manual selection and preventing information loss. These selected segments form a fundamental structure integrated into the GCN-based SOH estimation model, capturing inter-cycle dynamics and mitigating statistical distribution incongruities between offline training and online testing data. Validation with a widely accepted open-source dataset demonstrates that our method achieves precise SOH estimation, with a root mean squared error of less than 1$\%$.

\end{abstract}


\section{Introduction}
Lithium-ion batteries (LiBs) serve as a foundational technology for integrating intermittent renewable energy sources, which necessitate energy storage solutions to meet electrical demand \cite{CAMBOIM2024111091}. They are pivotal in combating climate change and promoting the transition to a decarbonized economy \cite{buchanan2024probabilistic}.  However, the performance of LiBs degrades over usage,  starting with slight degradation and accelerating after a certain quantity of charging and discharging processes. Therefore,  ensuring the accuracy of LiBs state of health (SOH) estimation is imperative to guarantee reliability,  performance, and longevity across various applications \cite{9320563}. 

The degradation of LiB's health results from internal electrochemical reactions that occur during charging and discharging,  where the reduction of active material and depletion of lithium inventory are the main factors contributing to their aging \cite{zhang2024state}.  Recently,  data-driven methods have gained prominence as noteworthy means of estimating the SOH of batteries,  owing to their inherent benefits of obviating the need for a comprehensive understanding of the intricate physical processes inherent in the batteries \cite{10018489}.  These techniques involve a series of iterative steps to enable machines to learn from data and make predictions, or estimations \cite{XIONG2022118915}. It also entails the selection of the most informative features from the data to ensure that the model can capture the complex relationships between the input and output features \cite{li2022nonlinear} \cite{9416833}. For instance, Wu $et\;al.$ \cite{WU2023109191} utilised the area of the constant-current charging and discharging voltage curves of LIBs as health features (HFs), and selected HFs of the battery and its SOH showed a strong correlation for SOH estimation. Lai $et\;al.$ \cite{LAI2023128971} converted the discharge voltage curves under dynamic working conditions into the trajectories corresponding to the constant current profiles and extracted the features from the reconstructed voltage curves by the Gaussian Process Regression model to estimate the SOH. Bamati $et\;al.$ \cite{9858904} fused the underlying feature from the discharge curve by adding exponential moving average as the health features and utilised nonlinear autoregressive with exogenous input to estimate SOH. Driscoll $et\;al.$ \cite{DRISCOLL2022104584} applied artificial neural networks to perform cycle SOH estimations by extracting features observed from patterns in the voltage, current, and temperature profiles during the charging process. Yao $et\;al.$ \cite{YAO2023106437} connected the health indicators extracted within the cycles from charging and discharging to graphic structure for estimating SOH using convolutional neural network (CNN),  long short-term memory (LSTM), and GraphSage. Wei $et\;al.$ \cite{WEI2023108947} predicted SOH by establishing a feature similarity-based graph from the charging cycle and devised a two-stage optimization model, encompassing a nonlinear integer optimization model integrated with a graph convolutional network (GCN) connected to LSTM networks equipped with dual attention mechanisms. 

In addition to the deliberate selection of architectural components and features, some researchers direct their attention toward the utilization of partial cycle segments in the construction of machine learning models. Tang $et\;al.$ \cite{TANG2023107734} employed voltage sampling points starting from 3.81V as input to their model, consisting of a CNN,  the LSTM,  and the convolutional block attention module, to estimate SOH.  Bockrath $et\;al.$  \cite{BOCKRATH2023120307} employed temporal convolutional network architecture to estimate battery SOH based on partial discharge profiles within different voltage ranges (4.20V to 3.70V, 3.70V to 3.50V, and 3.70V to 3.2V) corresponding to various state-of-charge (SOC) ranges. Kong $et\;al.$ \cite{KONG2021120114} extracted voltage-temperature health features from partial discharge voltage profiles and fed them into a Gaussian process regression-based model to estimate the SOH and RUL of lithium-ion batteries. Li $et\;al.$ \cite{LI2021116410} applied trial-and-error tests to select 225 data points from the charging segment and estimated battery capacity using CNN by transfer learning and network pruning techniques.  Lin $et\;al.$ \cite{LIN2022230774} employed a multi-feature-based multi-model fusion approach using voltage values, curve slope at fixed time intervals (200s-500s), temperature, and current features for SOH estimation.  Wang $et\;al.$ \cite{WANG2023120808} proposed a data aggregation and feature fusion scheme by graph neural networks (GNNs) to estimate the capacity of lithium-ion batteries by organizing the partial charging segment of voltage, current, and temperature in a graph structure. 

Despite the impressive progress achieved by the current research,  several challenges remain: 
\begin{itemize}
\item Partial discharge segments are relatively more attainable in real-world scenarios as opposed to complete cycles. Nonetheless, the arbitrary selection of such segments frequently obstructs their effectiveness, which in turn may result in the exclusion of crucial information.
\item The intricate process of feature engineering is computationally intensive and may encounter limitations when applied to testing data, particularly in scenarios involving battery aging dynamics.
\item Current SOH estimation approaches are typically trained based on features extracted from individual cycles.  This lack of consideration for inter-cycle degradation may lead to under-performance on test data that differs from the training data in terms of their data distributions.
\end{itemize}

To tackle the aforementioned challenges, an innovative methodology is presented to estimate the SOH by integrating statistical feature selection with GCN.  GCN has emerged as a prominent GNN to handle graph data \cite{APICELLA2023109867}. It can generate informative feature representations of network nodes by leveraging the graph convolutional layer to aggregate their neighbors \cite{REN2023126590}.  Such an approach can be instrumental in modeling the intricate relationships and interactions among battery cycles. The proposed approach employs the Matrix Profile (MP) algorithm \cite{7837898} to identify the most prominent voltage segment by exploring the temporal dynamics within each cycle.  Through the integration of inter-cycle degradation information with the selected voltage segment using a graph structure, and employing the GCN to process these features, the proposed methodology strives to surpass the performance of existing approaches.

The proposed methodology offers three key contributions to facilitate battery SOH estimation.
\begin{itemize}
\item Through the rigorous application of anomaly detection techniques to identify statistically significant partial segments, our method preserves critical information. This results in a substantial enhancement in the accuracy of SOH estimation.
\item By integrating voltage information directly into the graph structure, the proposed method eliminates the need for complex feature selection. This advancement not only enhances the model's adaptability but also significantly improves its efficiency and scalability. 
\item By capturing the temporal dynamics of deterioration across cycles, our method effectively addresses discrepancies in the statistical distribution between offline training and online testing data. This ensures greater reliability and robustness in SOH estimation, particularly in real-world scenarios.
\end{itemize}

The rest of the paper is organized into several sections.  In Section 2, the paper explains the experimental dataset used and the proposed GCN SOH estimation model.  Section 3 outlines the voltage segment selection process by MP and provides a comprehensive illustration of the online SOH estimation model. Section 4 presents the experimental outcomes.  The article concludes in Section 5, summarizing the key findings and implications derived from the paper.

\section{Preliminary}

\subsection{Experimental Dataset}
The widely used and publicly-accessible LiBs degradation dataset from MIT \cite{Attia2020} is used in this paper.  The commercial high-power APR18650M1A cells, utilizing LFP/graphite chemistry, have a 3.3 V nominal voltage and approximately 1.1Ah nominal capacity.  These batteries are connected to an Arbin LBT potentiostat housed in 48 channels under a temperature chamber with forced convection. The temperature of the chamber is maintained at 30$^{\circ}$C.  Charging is performed using one of 224 six-step protocols with the format "CC1-CC2-CC3-CC4" under 10-minute fast-charging conditions.  For the discharge process,  the current rate of 4C is applied, and the discharge is terminated when the voltage drops from 3.30V to 2.00V.

\begin{figure}[!htb]
\centering
\includegraphics[scale=0.45]{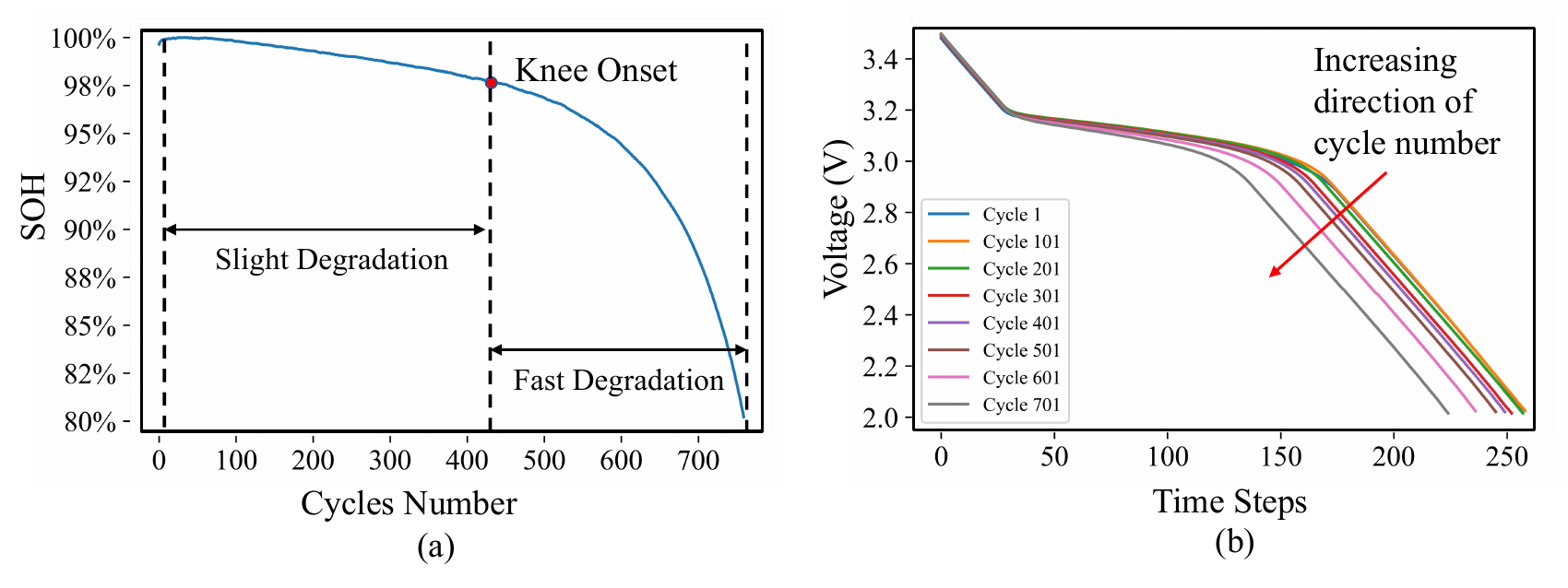}
\caption{(a): Lithium-ion battery SOH degradation of a randomly selected Battery \#2 from the benchmark dataset in \cite{Attia2020}, and (b): Different cycles discharge voltage of the Battery \#2. }
\label{MyFig1}
\end{figure}

The typical LiB SOH degradation path is shown in Fig. 1(a) using Battery \#2 from the dataset as an example.  In conjunction with the charging and discharging processes,  LiBs undergo two stages of degradation, initially experiencing slight degradation until reaching the knee onset \cite{10098615}, after which the deterioration starts accelerating. The end-of-life (EOL) of a battery is determined when its SOH declines to 80$\%$ of the initial capacity \cite{SON2022121712}.   Fig. 1(b) displays the discharge voltage curves for Battery \#2 when a constant discharge current is applied. A discernible trend is observed in the reduction of discharge voltage time intervals across successive cycles, which can be attributed to the declining SOH of the battery. 

However, owing to the presence of two distinct degradation stages, the voltage information utilized for model training during the slight degradation stage exhibits differential behavior compared to the data encountered during online testing stages. Consequently, this variance can lead to suboptimal model performance. Therefore, it becomes imperative to devise methodologies aimed at mitigating these discrepancies by incorporating degradation rate information during model training.  

\subsection{The Proposed GCN SOH estimation model}
As depicted in Fig. 2, the proposed GCN SOH estimation model is designed to capture the inter-cycle spatial information and provide insights into the overall degradation pattern of the battery utilized by GCN \cite{deng2021graph}. The cycles and their interconnections are depicted as a graph-structured $G(\mathcal{V},\mathcal{E})$.  Each cycle is represented as a node, where the voltage information is used as the node feature. The set of nodes is denoted by $\mathcal{V}$, where the node features are stored in the feature matrix $\mathbf X$. The correlations between the discharge voltage of individual cycles are utilized to construct interconnections between them, which are subsequently captured as edges in a graph and represented by $\mathcal{E}$.  The edge features $a_{ij}$ are stored at the adjacent matrix $\mathbf A$. 

\begin{figure}[hbt!]
\centering
\includegraphics[scale=0.45]{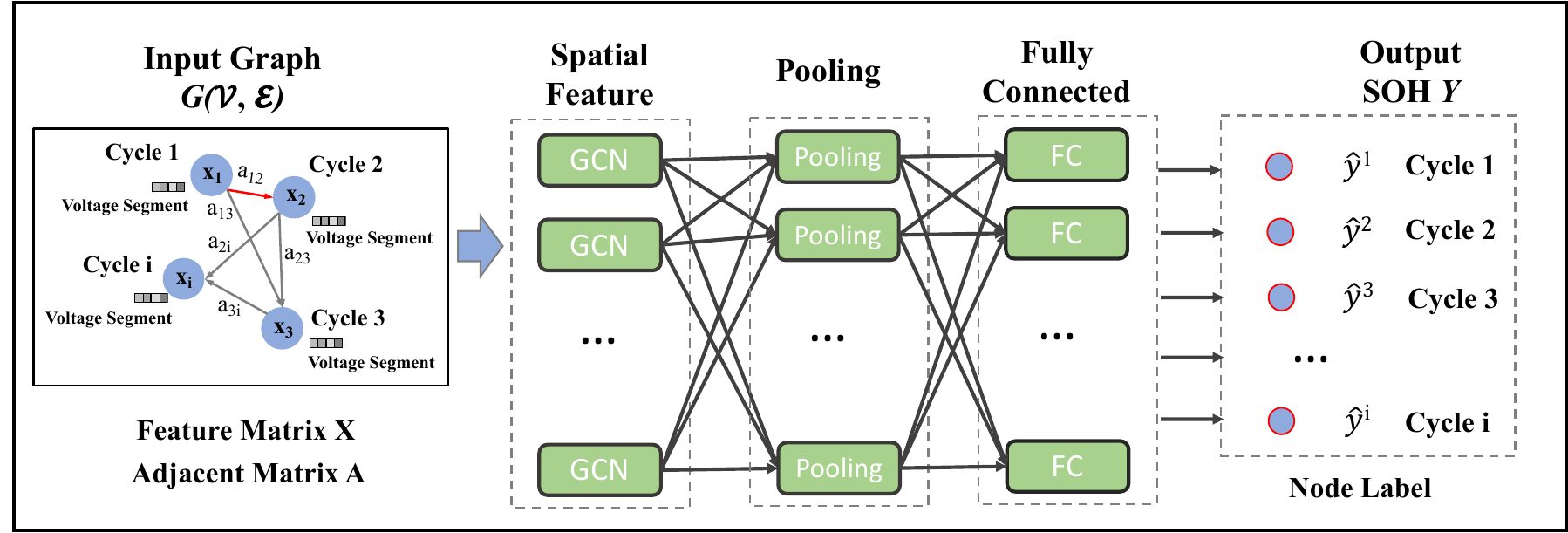}
\caption{Illustration of the GCN Estimation Model. }
\label{GCN}
\end{figure}

The initial feature matrix $\mathbf X$, formed by the voltage segment obtained from each node in the graph, is stored in a matrix $\mathbf H^{(0)}$. The graph convolution operation updates the node features by aggregating information from neighboring nodes. This is followed by a linear transformation using the weight matrix $\mathbf W^{(l)}$.  After the transformation,  a non-linear activation function $\boldsymbol \sigma(\boldsymbol \cdot)$ is applied, which introduces non-linearity into the model, allowing it to capture more complex relationships.  The final output of the GCN after $L$ layers is a set of node feature vectors $\mathbf H^{(l+1)}$. The final node features $\mathbf H^{(l+1)}$ is used for node classification, where each node is assigned a label as follows.
\begin{equation}
\begin{aligned}
&\mathbf H^{(0)}=\mathbf X\\
&\mathbf H^{(l+1)}=\boldsymbol \sigma (\mathbf {\hat D}^{-1/2} \mathbf A \mathbf {\hat D}^{-1/2} \mathbf H^{(l)} \mathbf W^{(l)})
\end{aligned}
\end{equation}
where $\mathbf X$ represents the feature matrix, $\mathbf A$ is the adjacent matrix, $\boldsymbol \sigma(\boldsymbol \cdot)$ is a non-linear activation function, and $\mathbf {\hat D}$ corresponds to the diagonal node degree matrix of $\mathbf A$, and $\mathbf W^{(l)}$ represents $l$-th layer's weight matrix.

Subsequently, the GCN layers are connected to pooling layers, which aggregate the node representations. Following the pooling layers, a fully connected layer is applied to these condensed node representations to generate the final SOH estimation for each node. 

\section{GCN-BASED SOH ESTIMATION USING PARTIAL \\ DISCHARGE CURVE}

\begin{figure}[!htb]
\centering
\includegraphics[scale=0.25]{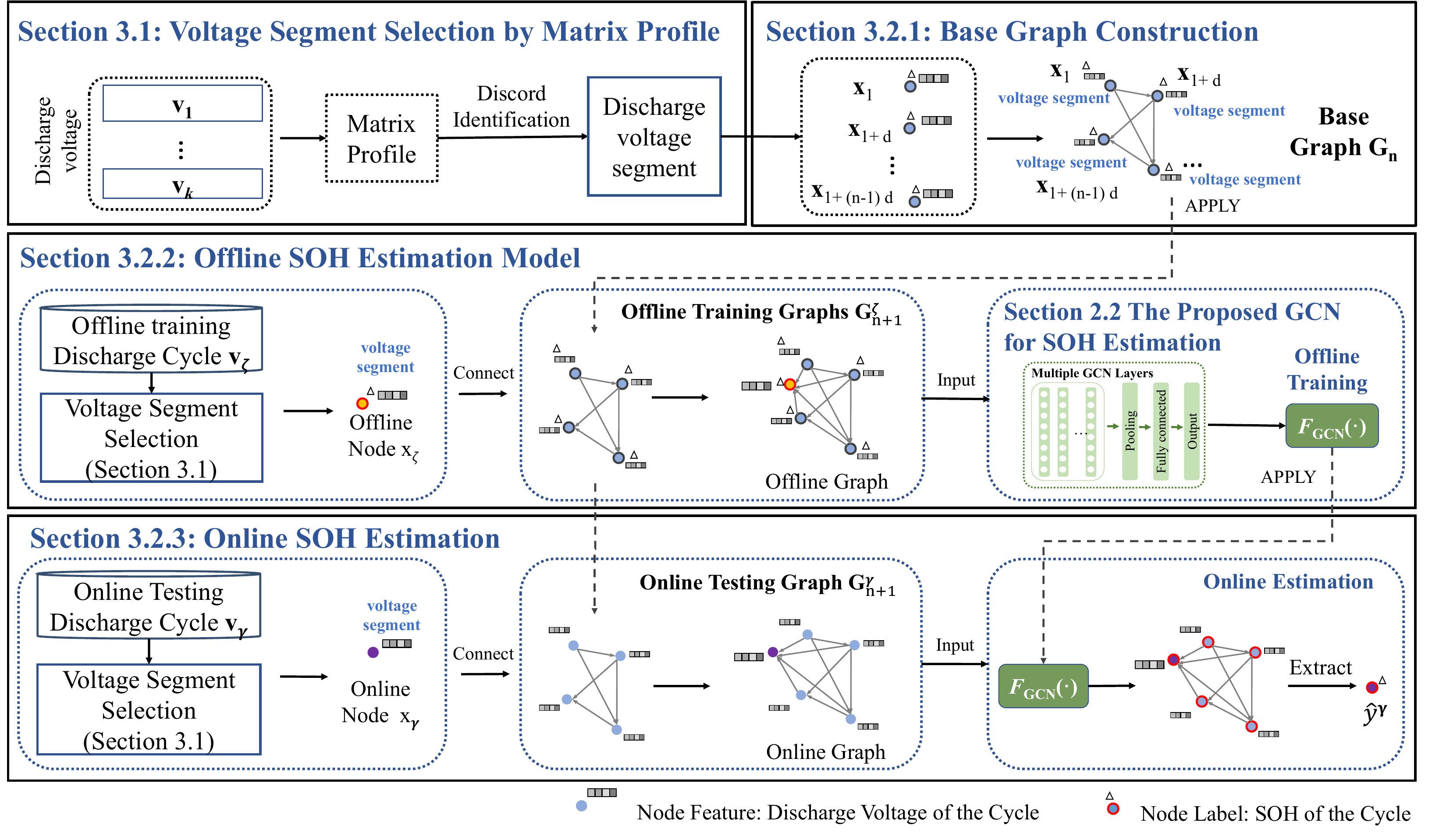}
\caption{GCN-based SOH Estimation framework using partial discharging cure}
\label{MyFig3}
\vspace{-4mm}
\end{figure}

To estimate the SOH of batteries online by GCN,  the process commences with voltage segment selection to identify the most prominent input segment. Subsequently, a base graph is constructed by amalgamating discharge voltage segments with identical intervals from the early degradation stage. Next, the voltage segment extracted from the offline training discharge cycle is connected to the base graph, thereby creating an offline training graph to train a GCN-based SOH estimation model for online estimation.  Finally,  the online estimation of SOH is facilitated by integrating the online discharge segment, identified through feature engineering, into the base graph. This amalgamation results in the creation of an online testing graph, which subsequently serves as the input for the proficient GCN-based SOH estimation model.  The overall process is illustrated in Fig. 3.

\subsection{Voltage Segment Selection by Matrix Profile}
The voltage information from each cycle is incorporated into the nodes of the graph. To identify the most salient segment from the other segments within the discharge voltage series, MP is employed.  MP is a data structure used in time series analysis to recognise and investigate patterns in sequential data \cite{8215529}.  Its wide range of applications in data mining encompasses the identification of trends, seasonality, and anomalies \cite{10027730}.  The fundamental component \textit{matrix profile} captures the minimum distance between the target subsequence and all other subsequences.  The minimum value in the \textit{matrix profile} indicates the repeated segment or motif,  whereas the maximum value represents the dissimilarity or anomaly \cite{8970797}. The steps to identify the most significant segment using MP are as follows.

Step 1: Calculate the \textit{matrix profile}. The first $k$ cycles of the LiB's early stage are selected. The $k$ cycles should be obtained during a period of slight degradation, where the battery health remains relatively consistent. Thus, the discharge voltage readings of these cycles $\mathbf v_{k} $ are joined as discharge time series $\mathbf{T}_k$:

\begin{equation}
\begin{aligned}
\mathbf T_k & =[ \mathbf v_1,\mathbf v_2,\ldots,\mathbf v_{k} ]\\
              & = [ v_1^{1},\ldots, v_{L_1}^{1},v_1^{2},\ldots,v_{L_2}^{2},\ldots , v_1^{k},\ldots,v_{L_k}^{k} ] \\
              & = [\dot v_1,\dot v_2,  \ldots, \dot v_\phi ]
\end{aligned}
\end{equation}
where the $\mathbf v_{k}$ is the discharge voltage of Cycle $k$,  $ {L_1}, {L_2},\ldots,{L_k} $ are the time steps of the cycles, $\phi={L_1}+{L_2}+\ldots+{L_k}$. 

By picking a query length of $m$ time steps,   $\mathbf T_k$ is partitioned into a series of overlapping segment $\mathbf T_{q,m}$,  which is called subsequence as follows:

\begin{equation}
\begin{aligned}
\mathbf T_{q,m}  =[\dot v_q, \dot v_{q+1}, \ldots, \dot v_{q+m-1}]
\end{aligned}
\end{equation}
where $1 \le q \le {\phi}-m+1$.

In time series analysis, the choice of subsequence length $m$ can be guided by the need to balance between capturing relevant patterns and maintaining computational efficiency.  it is recommended that the value of $m$ be a fraction of the length of the first cycle $L_1$ as $L_1/4 \leq m \leq L_1/2$.

For each subsequence $\mathbf{T}_{q,m}$, the Euclidean distances with all other subsequences are computed. The resulting smallest distance is recorded as $\dot{p}_q$ as shown in \eqref{eq:j}.  To prevent a subsequence from matching itself,  a size of $m/2$ exclusion zone is established:

\begin{equation}
\begin{aligned}
\dot p_q &= \min ( |\mathbf T_{q,m} - \mathbf T_{j,m}| )\label{eq:j}\\
\end{aligned}
\end{equation}
where $1\le j \le \phi-m+1$.

The minimum distance $\dot p_q$ for each subsequence is stored into \textit{matrix profile} $\mathbf P_k$ as shown in \eqref{eq:P_k}:
\begin{equation}
\begin{aligned}
\mathbf P_k&=[\dot p_1, \dot p_2,\ldots,\dot p_{\phi-m+1} ] \label{eq:P_k}
\end{aligned}
\end{equation}

The resulting $\textit{matrix profile}$ $\mathbf P_k$ is then partitioned into cycles that $\mathbf P_k = [\mathbf p_1, \mathbf p_2,\ldots,\mathbf p_{k} ]$, where $\mathbf p_{k}$ corresponds to the $\textit{matrix profile}$ of Cycle $k$. 

Step 2: Identify the discord. Cycle $2$ is chosen as the Golden Batch \cite{Yeh2019} to examine the discord.  The first $L$ time steps of the $\textit{matrix profile}$ $\mathbf p_{2}$  is denoted as $\mathbf p'_{2}= [p_1^{2},\ldots,p_{L}^{2}]$.  The segment with the highest peak $p_{\lambda} ^2$ at time step $\lambda$ is identified as the most distinguishable:
\begin{equation}
\begin{aligned}
p_{\lambda} ^2&= \max (\{ p_1^{2},\ldots,p_{L}^{2} \} )\\
\end{aligned}
\end{equation}

\begin{figure}[!htb]
\centering
\includegraphics[scale=0.45]{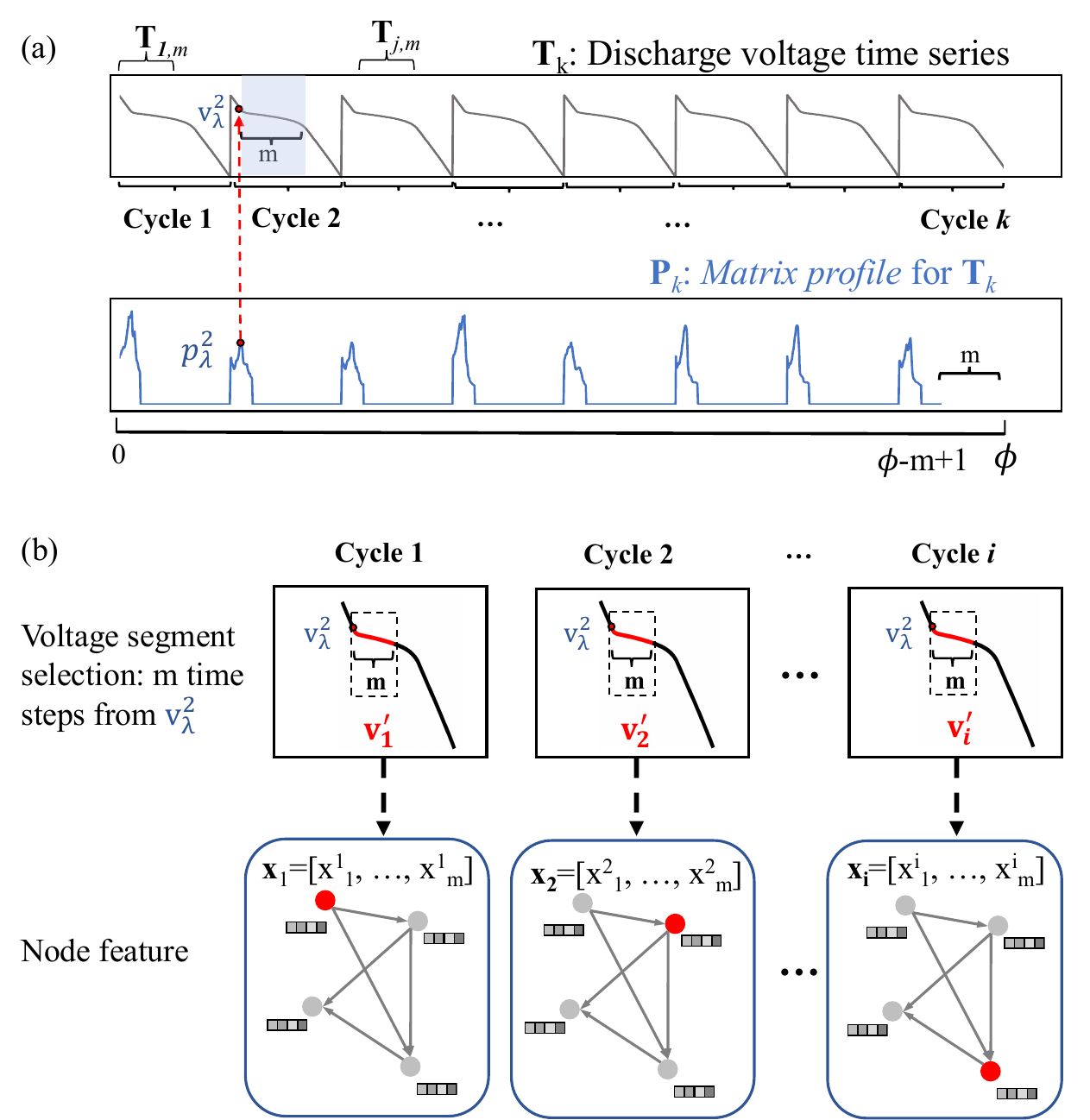}
\caption{(a) Discharge voltage time series and its \textit{matrix profile}; (b) Voltage segment selection of each cycle}
\label{MyFig4}
\end{figure}

By the highest peak $p_{\lambda}^2$,  the corresponding voltage value $v_{\lambda}^2$ is identified.  The voltage segment,  which begins from the voltage value $v_{\lambda}^2$ and spans a duration of $m$ time steps,  exhibits the highest degree of distinctiveness among all other voltage segments,  as illustrated in Fig. 4(a).  

Step 3: Select the voltage segment. Once the initial voltage reference $v_{\lambda}^2$ is established, it serves as the principal point of reference for determining the voltage segments for subsequent cycles.  When the voltage value $v_{\Theta}^i$ at time step $\Theta$ in Cycle $i$ is less than or equal to $v_{\lambda}^2$, the discharge voltage segment $\mathbf{v}'_i$ starting from time step $\Theta$ and comprising $m$ time steps is selected for Cycle $i$  as depicted in Fig. 4(b):
\begin{equation}
\begin{aligned}
\mathbf v'_i &=[v_\Theta^i, v_{\Theta+1}^i, \ldots, v_{\Theta +m}^i]
\end{aligned}
\end{equation}

The discharge voltage segment $\mathbf v'_i$ is utilized as the node feature and represented as $\mathbf x_i$ in the context of this study as follows:
\begin{equation}
\begin{aligned}
\mathbf x_i &=[x_1^i, x_2^i \ldots, x_{m}^i]
\end{aligned}
\end{equation}
where $\mathbf x^i $ is the node feature of Cycle $i$ by feature selection.

\subsection{Online SOH Estimation using GCN}
\subsubsection{Base Graph Construction}
Selecting $n$ cycles from the initial $k$ cycles, with a fixed interval of $d$ cycles between the cycles chosen, the base graph $G_{n}( \mathcal{V}_n, \mathcal{E}_n)$ is formed. The node feature $\mathbf x_i$,  identified by MP in Section 3.2, is incorporated into the feature matrix $\mathbf X_{n} \in \mathbb{R}^{n \times m}$.  The feature matrix $\mathbf X_{n}$ comprises a set of vectors that store the features associated with the $n$ nodes, where the features of the $i^{th}$ node are stored at the $i^{th}$ row of $\mathbf X_n$ as defined in \eqref{eq:X}:

\begin{equation}
\mathbf X_{n}=
\begin{bmatrix}
\mathbf x_1\\
\mathbf x_{1+d}\\
\vdots\\
\mathbf x_{1+(n-1)d}
\end{bmatrix}
=
\begin{bmatrix}
x_1^1&\ldots&x_{m}^1\\
x_1^{1+d}&\ldots&x_{m}^{1+d}\\
\ldots&\ldots&\ldots\\
x_1^{1+(n-1)d}&\ldots&x_{m}^{1+(n-1)d}
\end{bmatrix} \label{eq:X}
\end{equation}
where $m$ denotes the overall count of time steps present in the feature, and $d$ represents the interval between the selected cycles.

To denote the relationships between each pair of the $n$ nodes, the adjacent matrix $\mathbf A_{n} \in \mathbb{R}^{n \times n} $ is constructed by utilizing $a_{ij}$, which denotes the relationship between node $i$ and node $j$.   Owing to the progressive deterioration of LiBs over usage, the behavior of discharge cycles is influenced by those that precede them, thereby rendering the relationship among discharge cycles in the graph to be directed.  It is noteworthy that when $i=j$, the influence of a discharge cycle on itself is denoted by $a_{ij}=1$. Hence, the adjacency matrix $\mathbf{A}_{n}$ takes the form of a right triangular matrix, as illustrated below:

\begin{equation}
\mathbf A_{n}=
\left[\begin{array}{ccccc}
1&a_{12}&\cdots&\cdots&a_{1n}\\
0&1&a_{23}&\cdots&a_{2n}\\
\vdots&\ddots&\ddots&\ddots&\vdots\\
\vdots&\ddots&\ddots&\ddots&a_{(n-1)n}\\
0&\cdots&\cdots&0&1\\
\end{array}\right]
\end{equation}
where $n$ is the total count of notes.

The value of $a_{ij}$ is computed using the Pearson correlation coefficient, a statistical metric that gauges the magnitude and direction of the linear association between two variables as follows.

\begin{equation}
\begin{aligned}
a_{ij}=\frac{\mathbb{E} [\mathbf x_i \mathbf x_j]-\mathbb{E} [\mathbf x_i]\mathbb{E}[\mathbf x_j]}{\sqrt{\mathbb{E} [\mathbf x_i^2]-\mathbb{E} [\mathbf x_i]^2}\sqrt{\mathbb{E}[\mathbf x_j^2]-\mathbb{E} [\mathbf x_j]^2}}
\end{aligned}
\end{equation}
where $\mathbb{E} [\boldsymbol \cdot]$ is the expectation of variables,  $i \in [1,n]$,  $j \in [1,n]$.

This coefficient is bounded within the range of -1 to 1, where $a_{ij}= 1$ indicates a perfect positive linear correlation, meaning that as one variable increases, the other also increases proportionally and $a_{ij}= 0$ suggests no linear correlation between the variables, meaning that they are statistically independent. In the context of battery cycles, as the battery undergoes degradation, the similarity between these voltage profiles and those from earlier cycles during the slight degradation stage decreases. This results in lower Pearson correlation coefficient values when calculated between these cycles. Thus, it offers a mechanism to assess the strength of the link between cycles based on the similarity of their respective features.

The node labels $\mathbf y_{n}$, compiling the measured SOH for $n$ cycles, are represented as follows:

\begin{equation}
\begin{aligned}
\mathbf y_{n} = [y^1,   y^2,  \ldots,  y^{n}]^T
\end{aligned}
\end{equation}
where $y^{n}$ is the measured SOH of the $n^{th}$ cycle in base graph $G_{n}$.

During the process of constructing the base graph using a predetermined set of cycles and employing the Pearson correlation coefficient, it is noteworthy that this metric remains invariant. This steadfastness in measurement serves as a guarantor of both consistency and stability within the base graph. Consequently, the base graph is established as a stable structural framework, poised to facilitate connections with other cycles effectively.

\subsubsection{Offline SOH Estimation Model}
Subsequent to Cycle $k$,  each training discharge Cycle $\zeta$ undergoes the same segment extraction procedure in Section 3.1 to generate the node feature $\mathbf x_{\zeta}$. It is combined with the base graph $G_{n}$ from Section 3.2.1 to create the offline training graph $G_{n+1}^{\zeta}$. The feature matrix $\mathbf X_{n+1}^\zeta \in \mathbb{R}^{(n+1) \times m} $ and the adjacent matrix $\mathbf A_{n+1}^\zeta \in \mathbb{R}^{(n+1) \times (n+1)} $ are defined as follows:

\begin{equation}
\mathbf X_{n+1}^{\zeta}=
\begin{bmatrix}
\mathbf X_{n}\\
\mathbf x_{\zeta}
\end{bmatrix}
=
\begin{bmatrix}
x_1^1&\ldots&x_{m}^1\\
x_1^{1+d}&\ldots&x_{m}^{1+d}\\
\ldots&\ldots&\ldots\\
x_1^{1+(n-1)d}&\ldots&x_{m}^{1+(n-1)d}\\
x_1^{\zeta}&\ldots&x_{m}^{\zeta}
\end{bmatrix}
\end{equation}
\begin{equation}
\mathbf A_{n+1}^{\zeta}=
\left[\begin{array}{cccccc}
1&a_{12}&\cdots&\cdots&a_{1n}&a_{1\zeta}\\
0&1&a_{23}&\cdots&a_{2n}&a_{2\zeta}\\
\vdots&\ddots&\ddots&\ddots&\vdots&\vdots\\
\vdots&\ddots&\ddots&\ddots&a_{(n-1)n}&\vdots\\
0&\cdots&\cdots&0&1&a_{n\zeta}\\
0&\cdots&\cdots&\cdots&0&1\\
\end{array}\right]
\end{equation}
where  $a_{n\zeta}$ is the correlation coefficient between Cycle $n$ and Cycle $\zeta$, $\zeta \in[k,K_{tr}]$,  $K_{tr}$ is the end of the training cycles.

The offline training feature matrix $\mathbf X_{n+1}^\zeta$ and adjacent matrix $\mathbf A_{n+1}^\zeta$ are the input to train the SOH estimation model as described in Section 2.2. Node-level learning is employed, where the model predicts the SOH for each node and outputs a vector $\mathbf {\hat{y}}_{n+1}^\zeta$ containing these predicted values.  During training, the model minimizes the difference in the predicted SOH $\mathbf {\hat{y}}_{n+1}^\zeta$ in relation to the measured SOH $\mathbf y_{n+1}^\zeta=[y^1, y^2, \ldots, y^n, y^{\zeta}]^T$ as follows:

\begin{equation}
\begin{aligned}
&\min\sum{(\mathbf y_{n+1}^\zeta-\mathbf {\hat{y}}_{n+1}^\zeta)}^2\\
\end{aligned}
\end{equation}
\begin{equation}
\begin{aligned}
&\mathbf {s.t. \quad \hat{y}}_{n+1}^\zeta=
\begin{bmatrix}
 \hat y^1\\
\hat y^2\\
\vdots\\
\hat y^{n}\\
\hat y^{\zeta}
\end{bmatrix}
=\emph{F}_{GCN}(\mathbf X_{n+1}^\zeta,  \mathbf A_{n+1}^\zeta)
\end{aligned}
\end{equation}
where $\mathbf {\hat{y}}_{n+1}^\zeta$ is the estimated SOH for all the nodes in the offline training graph,  $\emph{F}_{GCN}(\boldsymbol \cdot)$ stands for the GCN-based SOH estimation model.

\subsubsection{Online SOH Estimation}
The node feature $\mathbf x_{\gamma}$ for the online discharge Cycle $\gamma$ is obtained using the methodology described in Section 3.1. The online graph $G_{n+1}^{\gamma}$ is then constructed by combining $\mathbf x_{\gamma}$ with the base graph $G_n$ from Section 3.2.1. Consequently, the online feature matrix $\mathbf X_{n+1}^\gamma$ and online adjacent matrix $\mathbf A_{n+1}^\gamma$  are defined as follows:
\begin{equation}
\mathbf X_{n+1}^\gamma=
\begin{bmatrix}
\mathbf X_{n}\\
\mathbf x_{\gamma}
\end{bmatrix}
=
\begin{bmatrix}
x_1^1&\ldots&x_{m}^1\\
x_1^{1+d}&\ldots&x_{m}^{1+d}\\
\ldots&\ldots&\ldots\\
x_1^{1+(n-1)d}&\ldots&x_{m}^{1+(n-1)d}\\
x_1^{\gamma}&\ldots&x_{m}^{\gamma}
\end{bmatrix}
\end{equation}

\begin{equation}
\mathbf A_{n+1}^\gamma=
\left[\begin{array}{cccccc}
1&a_{12}&\cdots&\cdots&a_{1n}&a_{1\gamma}\\
0&1&a_{23}&\cdots&a_{2n}&a_{2\gamma}\\
\vdots&\ddots&\ddots&\ddots&\vdots&\vdots\\
\vdots&\ddots&\ddots&\ddots&a_{(n-1)n}&\vdots\\
0&\cdots&\cdots&0&1&a_{n\gamma}\\
0&\cdots&\cdots&\cdots&0&1\\
\end{array}\right]
\end{equation}
where $a_{n\gamma}$ is the correlation coefficient between Cycle $n$ and Cycle $\gamma$.

The online feature matrix $\mathbf X_{n+1}^\gamma$ and adjacent matrix $\mathbf A_{n+1}^\gamma$ are processed by the trained SOH estimation model $\emph{F}_{GCN}(\boldsymbol \cdot)$ to output node labels $\mathbf {\hat{y}}_{n+1}^\gamma$ for the online graph.  Accordingly,  the estimated SOH $\hat y^{\gamma}$ of the online Cycle $\gamma$ is obtained as follows:

\begin{equation}
\begin{aligned}
\mathbf {\hat{y}}_{n+1}^\gamma=
\begin{bmatrix}
\hat y^1\\
\hat y^2\\
\vdots\\
\hat y^{n}\\
\hat y^\gamma
\end{bmatrix}
=\emph{F}_{GCN}(\mathbf X_{n+1}^\gamma,  \mathbf A_{n+1}^\gamma)
\end{aligned}
\end{equation}
where $\mathbf {\hat{y}}_{n+1}^\gamma$ is the estimated SOH for all nodes in the online graph,  $\hat y^{\gamma}$ is the estimated SOH for Cycle $\gamma$.

\section{EXPERIMENT RESULT AND DISCUSSION}

Four batteries from the dataset, namely Battery \#2, Battery \#6, Battery \#16, and Battery \#25,  selected from different charging channels are utilized as shown in Table 1.  The batteries exhibit distinct operational lifetimes, with Battery \#2 completing 760 cycles until reaching its EOL,  Battery \#6 completing 732 cycles, Battery \#16 completing 930 cycles, and Battery \#25 completing 467 cycles.

\begin{table}[!htb]
	\scriptsize
	\renewcommand{\arraystretch}{1.2}
	\caption{Batteries Specification  \cite{Attia2020}} 
	\vspace{-4mm}
	\label{table_1}
	\begin{center}
{\begin{tabular}{p{0.15\textwidth}>{\centering\arraybackslash}m{0.1\textwidth}>{\centering\arraybackslash}m{0.1\textwidth}}
\hline\hline
\\[-1em]
 				\multicolumn{1}{c} {\textbf {Charge Protocol}} & \multicolumn{1} {c}{\textbf{Battery Number}}& \textbf{Life Cycle} \\
\hline 
\multicolumn{1}{c} {3.6-6.0-5.6-4.755C} & 2&760 \\
\multicolumn{1}{c} {8.0-4.4-4.4-3.940C} & 6&732 \\
\multicolumn{1}{c} {7.0-4.8-4.8-3.652C} & 16&930 \\
\multicolumn{1}{c} {8.0-7.0-5.2-2.680C} & 25& 467\\
\hline
\hline
			\end{tabular}}
	\end{center}
\vspace{-4mm}
\end{table}

\subsection{Voltage Segment Selection by Matrix Profile}
In accordance with the methodology outlined in Section 3.1, Battery \#2 is selected to demonstrate the process of voltage segment selection using MP. The first $100$ discharge voltage cycles were concatenated, and their $\textit{matrix profile}$ was generated.  For brevity, only the first $3$ cycles are presented in Fig. 5(a). The discord in the second cycle was identified by its $\textit{matrix profile}$ at a voltage value of $3.25V$.  Consequently,  100 time steps from voltage 3.25V are selected as the voltage segment as the input for SOH estimation for Battery 2 as illustrated in Fig. 5(b).  This same approach is applied to Batteries \#6, \#16, and \#25, where discords were identified at voltages of $3.18V$, $3.22V$, and $3.24V$, respectively.

\begin{figure}[!htb]
\centering
\includegraphics[scale=0.4]{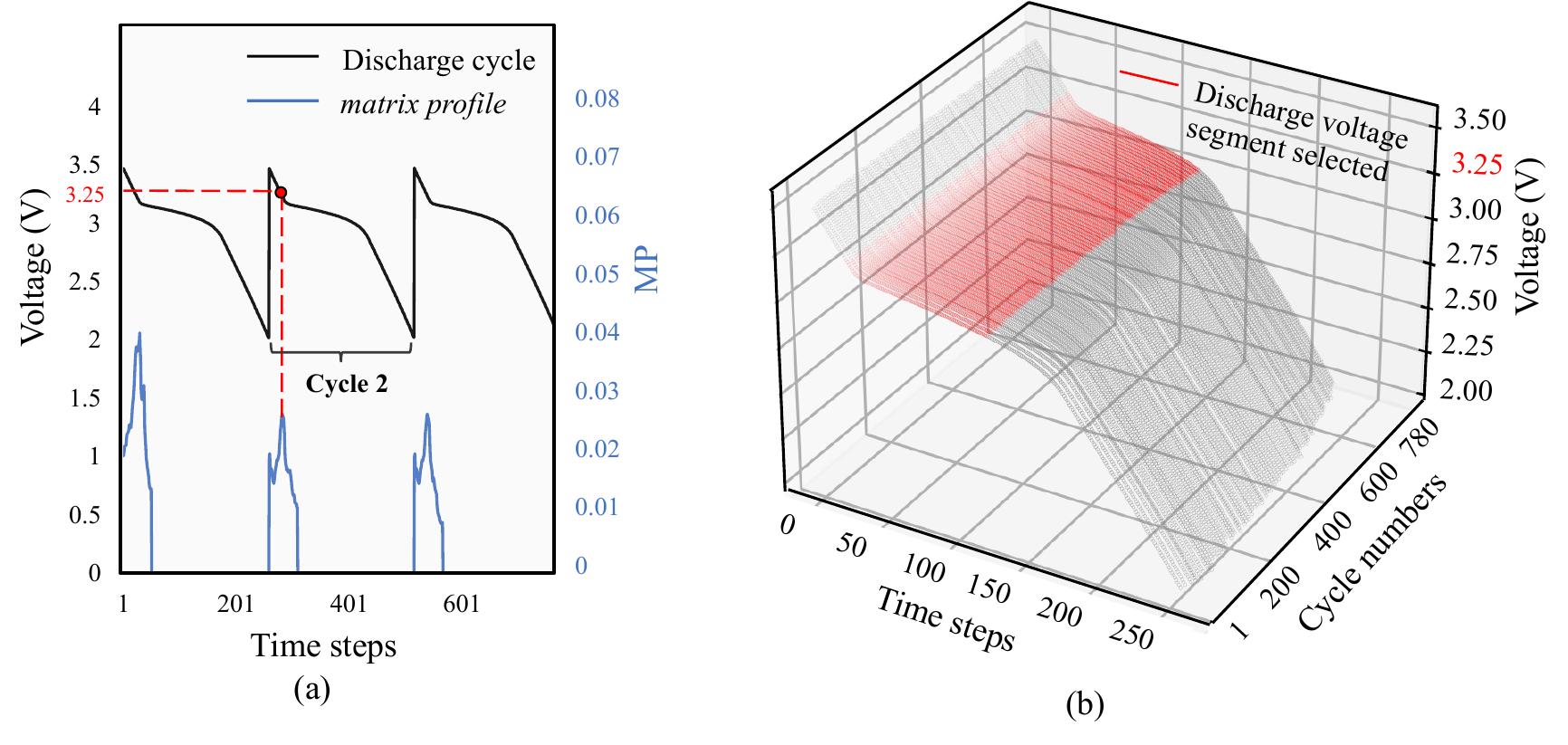}
\caption{ (a): Discharge voltage time series and its \textit{matrix profile} of Battery \#2 [17], and (b):  Discharge voltage segment of 100 time steps from $3.25V$ of Battery \#2 [17]. }
\label{fig4}
\end{figure}

\subsection{Online SOH Estimation by GCN}
\subsubsection{Base Graph Construction}
To form the base graph $G_{10}( \mathcal{V}_{10}, \mathcal{E}_{10})$,  ten nodes are selected from the first one hundred cycles of each battery, with nodes being chosen at intervals of ten cycles, specifically at the $1^{st}$,  $11^{th}$,  $\cdots$, $91^{st}$  cycles. The formation of the feature matrix $\mathbf{X}_{10}$ and the adjacent matrix $\mathbf{A}_{10}$ is carried out for each battery in relation to the constructed base graph.

\subsubsection{Offline SOH Estimation Model}
The model for estimating SOH is trained using the partial discharging segment from the initial $70\%$ data,  spanning from Cycle 101 to EOL.  For each cycle in this training data, the base graph $G_{10}$ is combined to form a training graph $G_{11}$ consisting of $11$ nodes. The corresponding feature matrix $\mathbf{X_{11}}$ and adjacent matrix $\mathbf{A_{11}}$ are then used as inputs for the SOH estimation model using GCN.  It consists of a single GCN layer with 128 neurons, followed by a global attention-pooling layer.  Before the output layer, an additional normal layer consisting of 300 neurons is incorporated. To minimize errors, the optimizer Adam is utilized, and the model is trained over $200,000$ epochs to determine the optimal weights. The training is executed in a single desktop with AMD Ryzen 9 5950X, 16-core Processor, CPU @3.4GHz,  128 GB RAM with Nvidia Geforce RTX 3080Ti GPU. 

\subsubsection{Online SOH Estimation}
Upon completion of the training phase, the remaining $30\%$ of the cycles are reserved as online cycles for testing purposes.  Specifically,  Battery \#2, Battery \#6, Battery \#16, and Battery \#25, undergo 198, 189, 249, and 110 cycles, respectively.  The proposed methods are evaluated using a pair of performance measures, namely the mean absolute error (MAE) and root mean squared error (RMSE) as follows:

\begin{equation}
\begin{aligned}
\mathbf {MAE}=\frac{1}{\Gamma}\sum_{\gamma=1}^{\Gamma}{\bigg|y^\gamma -\hat y^{\gamma}\bigg|}\\
\mathbf {RMSE}=\sqrt{\frac{1}{\Gamma}\sum_{\gamma=1}^{\Gamma}{(y^\gamma -\hat y^{\gamma})^2}}
\end{aligned}
\end{equation}
where $\Gamma$ is the total number of online cycles,  $\hat y^{\gamma}$ denotes the estimated SOH,  and $y^\gamma$ corresponds the measured SOH.

The SOH estimation results for Battery \#2, Battery \#6, Battery \#16, and Battery \#25 achieved RMSE values of 0.0087, 0.0082, 0.0089 and 0.0063, respectively.  To verify the effectiveness of the voltage segment selected by MP.  Six discrete voltage segments of $100$ time steps with different starting voltages separated by intervals of $0.02V$ are used as input to train and test the GCN estimation.  The results indicate voltage segments before $3.18V$ are getting estimation results of RMSE lower than $0.0025$. In particular, the partial segment from our proposed methods achieves the best performance among the different voltage segments,  as shown in Fig. 6 and summarised in Table 2.   This observation emphasizes the importance of selecting an appropriate voltage segment to enhance the estimation performance.

\begin{figure}[!htb]
\centering
\includegraphics[scale=0.45]{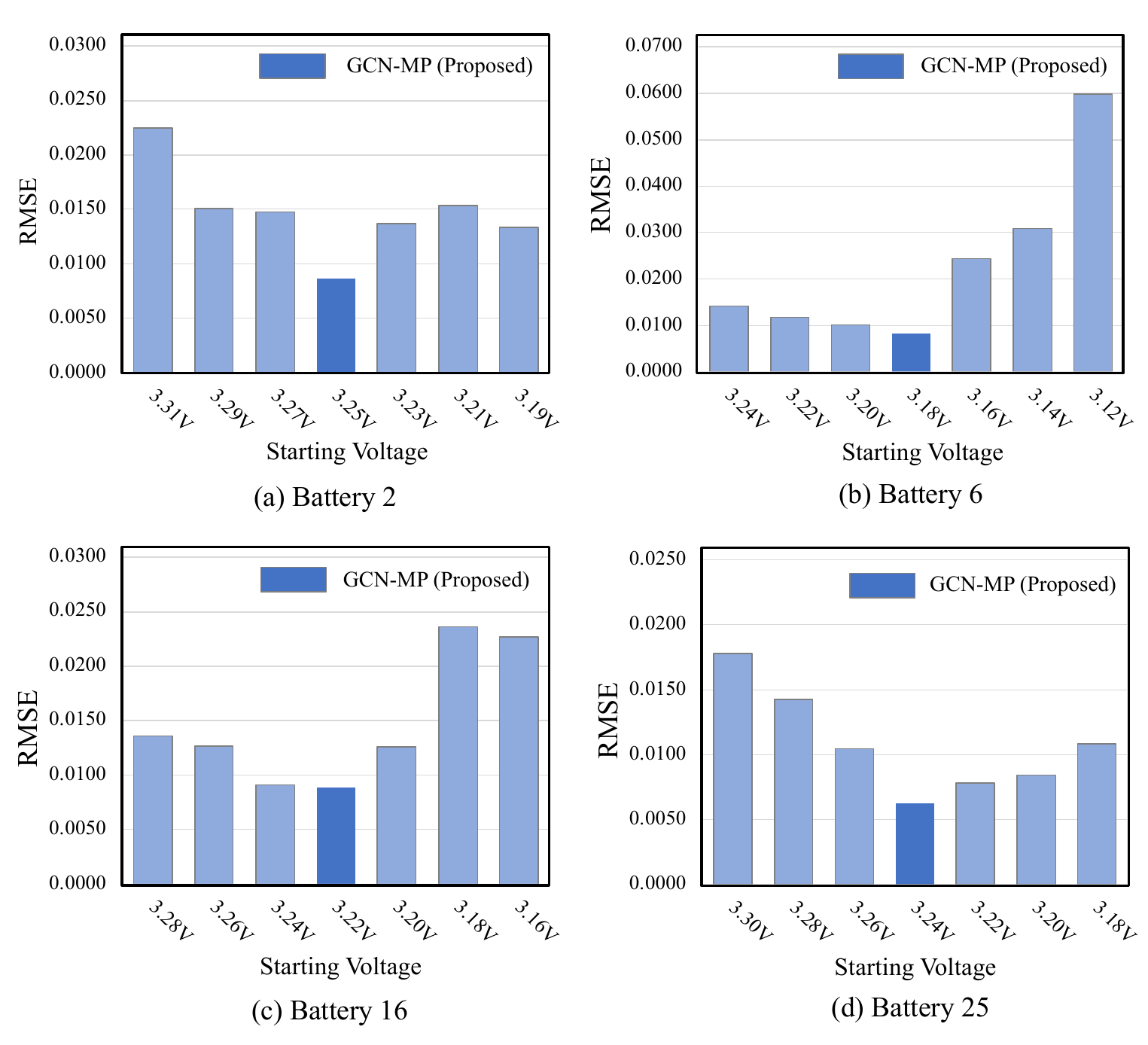}
\caption{RMSE comparison of different voltage segments by GCN for (a): Battery 2; (b): Battery 6; (c): Battery 16; (d): Battery 25.}
\label{diffvol}
\end{figure}

\begin{table*}[hbt!]
	\scriptsize
	\renewcommand{\arraystretch}{1.3}
\caption{RMSE comparisons of the proposed method and different voltage segments}
\begin{center}
\begin{tabular}{>{\centering\arraybackslash}m{0.08\textwidth}>{\centering\arraybackslash}m{0.08\textwidth}>{\centering\arraybackslash}m{0.08\textwidth}>{\centering\arraybackslash}m{0.08\textwidth}>{\centering\arraybackslash}m{0.08\textwidth}>{\centering\arraybackslash}m{0.08\textwidth}>{\centering\arraybackslash}m{0.08\textwidth}>{\centering\arraybackslash}m{0.08\textwidth}}
\hline
\hline
\\[-1.5em]
\multicolumn{2}{c}{Battery \#2} & \multicolumn{2}{c}{Battery \#6} & \multicolumn{2}{c}{Battery \#16} & \multicolumn{2}{c}{Battery \#25} \\
\hline
\\[-1.5em]
$v$    & RMSE      & $v$     & RMSE      & $v$     & RMSE       & $v$     & RMSE       \\
\\[-1.5em]
\hline
\\[-1.5em]
3.31V & 0.0225    & 3.24V & 0.0143    & 3.28V & 0.0136     & 3.30V & 0.0178     \\
\\[-1.5em]
3.29V & 0.0151    & 3.22V & 0.0117    & 3.26V & 0.0127     & 3.28V & 0.0143     \\
\\[-1.5em]
3.27V & 0.0147    & 3.20V & 0.0101    & 3.24V & 0.0092     & 3.26V & 0.0105     \\
\\[-1.5em]
\textbf{3.25V*} & \textbf{0.0087}    & \textbf{3.18V*} & \textbf{0.0082}    & \textbf{3.22V*} & \textbf{0.0089}     & \textbf{3.24V*} & \textbf{0.0063}     \\
\\[-1.5em]
3.23V & 0.0137    & 3.16V & 0.0244    & 3.20V & 0.0126     & 3.22V & 0.0078     \\
\\[-1.5em]
3.21V & 0.0154    & 3.14V & 0.0308    & 3.18V & 0.0236     & 3.20V & 0.0084     \\
\\[-1.5em]
3.19V & 0.0133    & 3.12V & 0.0599    & 3.16V & 0.0227     & 3.18V & 0.0109    \\
\hline
\hline
\multicolumn{2}{l}{* Proposed Method}
\end{tabular}
\end{center}
\end{table*}

Furthermore, the proposed method's efficacy is evaluated by comparing it to the widely-used non-GNN algorithm LSTM, which solely relies on temporal information within the cycles. Two feature selection techniques,  LSTM-MP and LSTM-DTW, were applied.  LSTM-MP employed the same voltage segment as the proposed method, which consisted of 100 time steps from the discord identified by MP,  whereas LSTM-DTW employed the complete discharge voltage data and synchronized them with the first discharge voltage cycle, as described in \cite{9589504}.  The LSTM architecture comprises two layers of LSTMs with 200 and 300 neurons each followed by a fully connected layer consisting of 100 neurons and 300 epochs are run. The resulting estimations of the batteries and the distribution of the error of the three methods are illustrated in Fig. 7. The proposed GCN-MP method, which incorporates inter-cycle degradation information through the graph structure, outperformed the other two methods.  In contrast, the sequential algorithm fails to capture such information, leading to suboptimal estimation results. Table 3 presents a comprehensive overview of the RMSE and MAE results for the four batteries.

\begin{figure*}[!htb]
\centering
\includegraphics[scale=0.4]{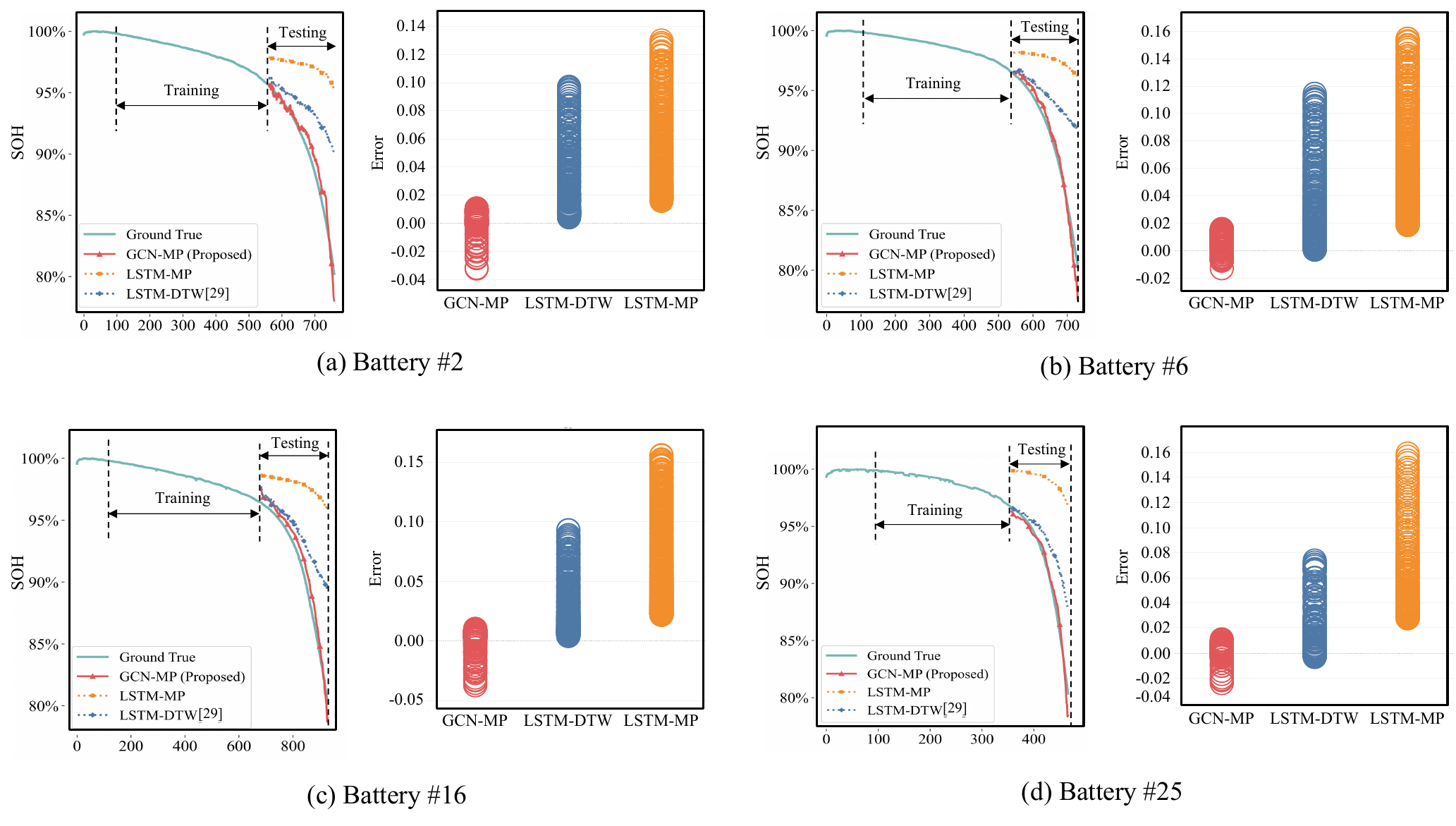}
\caption{SOH estimation result comparisons of three methods for (a) Battery \#2; (b) Battery \#6; (c) Battery \#16; (d) Battery \#25}
\label{soh_result}
\end{figure*}

\begin{table*}[hbt!]
	\scriptsize
	\renewcommand{\arraystretch}{1.3}
\caption{Comparative analysis between the proposed method and alternative approaches by index RMSE and MAE}
\begin{tabular}{>{\centering\arraybackslash}m{0.22\textwidth}>{\centering\arraybackslash}m{0.06\textwidth}>{\centering\arraybackslash}m{0.06\textwidth}>{\centering\arraybackslash}m{0.06\textwidth}>{\centering\arraybackslash}m{0.06\textwidth}>{\centering\arraybackslash}m{0.06\textwidth}>{\centering\arraybackslash}m{0.06\textwidth}>{\centering\arraybackslash}m{0.06\textwidth}>{\centering\arraybackslash}m{0.06\textwidth}}\hline
\hline
\\[-1em]
  \multirow{2}{*}{\textbf{Method Name}} & \multicolumn{2}{c}{Battery \#2}  & \multicolumn{2}{c}{Battery \#6}       & \multicolumn{2}{c}{Battery \#16}  &\multicolumn{2}{c}{Battery \#25}      \\\cline{2-9}
\\[-1em]
           & RMSE      & MAE    & RMSE      & MAE    & RMSE       & MAE    & RMSE       & MAE    \\\hline
\\[-1em]
\textbf{GCN-MP   (Proposed) }& \textbf{0.0087 }   & \textbf{0.0067} & \textbf{0.0082}    & \textbf{0.0065} & \textbf{0.0089}     & \textbf{0.0082} & \textbf{0.0063}     &\textbf{0.0049} \\\hline
\\[-1em]
LSTM-DTW [30]  & 0.0436    & 0.0344 & 0.0475    & 0.0356 & 0.0367     & 0.0277 & 0.0325     & 0.0233 \\\hline
\\[-1em]
LSTM-MP              & 0.0769   & 0.0677 & 0.0776    & 0.0662 & 0.0770     & 0.0660 & 0.0841     & 0.0746 \\\hline
\hline
\end{tabular}
\vspace{-2mm}
\end{table*}

\section{Conclusion}
The acquisition of relevant voltage segments and the alignment of statistical distributions between the training and testing datasets are critical factors in improving the performance of data-driven methods.  This study introduces a novel methodology to harness the spatio-temporal degradation dynamics inherent in battery cycles by employing a graph convolutional network (GCN) for battery state of health (SOH) estimation. Notably, the accuracy of SOH estimation is further augmented through the implementation of a systematic statistical approach for the selection of voltage segments. The most salient partial discharge voltage segment is identified statistically by the Matrix Profile and its effectiveness on SOH estimation is evaluated by comparison with manual and arbitrarily selected voltage segments.  The outcomes validate the effectiveness of the proposed method in achieving precise SOH estimation,  underscoring the importance of voltage segment selection in data-driven approaches to SOH estimation.  Furthermore, integrating the identified voltage segment into the graph structure enables the capture of degradation correlation between cycles, which is utilized by the graph convolutional network for online SOH estimation. The proposed method outperforms the approaches solely reliant on inner-cycle information by effectively capturing the inter-cycle dynamics, resulting in accurate SOH estimation. In future works, it is worthwhile to explore employing transfer learning techniques, where the estimation model derived from established battery datasets is applied to online batteries to enhance the efficiency and flexibility of real-time battery health estimations. In addition, the refinement of models to handle diverse operational parameters, including variable temperature and discharge rates, and incomplete charging scenarios, is needed to improve adaptability in real-world applications.

\bibliographystyle{IEEEtran}
\bibliography{Reference.bib}

\begin{thebibliography}{10}
\providecommand{\url}[1]{#1}
\csname url@samestyle\endcsname
\providecommand{\newblock}{\relax}
\providecommand{\bibinfo}[2]{#2}
\providecommand{\BIBentrySTDinterwordspacing}{\spaceskip=0pt\relax}
\providecommand{\BIBentryALTinterwordstretchfactor}{4}
\providecommand{\BIBentryALTinterwordspacing}{\spaceskip=\fontdimen2\font plus
\BIBentryALTinterwordstretchfactor\fontdimen3\font minus
  \fontdimen4\font\relax}
\providecommand{\BIBforeignlanguage}[2]{{%
\expandafter\ifx\csname l@#1\endcsname\relax
\typeout{** WARNING: IEEEtran.bst: No hyphenation pattern has been}%
\typeout{** loaded for the language `#1'. Using the pattern for}%
\typeout{** the default language instead.}%
\else
\language=\csname l@#1\endcsname
\fi
#2}}
\providecommand{\BIBdecl}{\relax}
\BIBdecl

\bibitem{CAMBOIM2024111091}
M.~M. Camboim and M.~Giesbrecht, ``Online state of health estimation of
  lithium-ion batteries through subspace system identification methods,''
  \emph{Journal of Energy Storage}, vol.~85, p. 111091, 2024.

\bibitem{buchanan2024probabilistic}
S.~Buchanan and C.~Crawford, ``Probabilistic lithium-ion battery
  state-of-health prediction using convolutional neural networks and gaussian
  process regression,'' \emph{Journal of Energy Storage}, vol.~76, p. 109799,
  2024.

\bibitem{9320563}
Y.~Qin, S.~Adams, and C.~Yuen, ``Transfer learning-based state of charge
  estimation for lithium-ion battery at varying ambient temperatures,''
  \emph{IEEE Transactions on Industrial Informatics}, vol.~17, no.~11, pp.
  7304--7315, 2021.

\bibitem{zhang2024state}
B.~Zhang, W.~Liu, Y.~Cai, Z.~Zhou, L.~Wang, Q.~Liao, Z.~Fu, and Z.~Cheng,
  ``State of health prediction of lithium-ion batteries using particle swarm
  optimization with levy flight and generalized opposition-based learning,''
  \emph{Journal of Energy Storage}, vol.~84, p. 110816, 2024.

\bibitem{10018489}
Y.~Qin, A.~Arunan, and C.~Yuen, ``Digital twin for real-time li-ion battery
  state of health estimation with partially discharged cycling data,''
  \emph{IEEE Transactions on Industrial Informatics}, pp. 1--11., 2023.

\bibitem{XIONG2022118915}
R.~Xiong, J.~Huang, Y.~Duan, and W.~Shen, ``Enhanced lithium-ion battery model
  considering critical surface charge behavior,'' \emph{Applied Energy}, vol.
  314, p. 118915, 2022.

\bibitem{li2022nonlinear}
W.~Li, C.~Yang, and S.~E. Jabari, ``Nonlinear traffic prediction as a matrix
  completion problem with ensemble learning,'' \emph{Transportation Science},
  vol.~56, no.~1, p. 52–78, jan 2022.

\bibitem{9416833}
Y.~Qin, W.-T. Li, C.~Yuen, W.~Tushar, and T.~K. Saha, ``Iiot-enabled health
  monitoring for integrated heat pump system using mixture slow feature
  analysis,'' \emph{IEEE Transactions on Industrial Informatics}, vol.~18,
  no.~7, pp. 4725--4736, 2022.

\bibitem{WU2023109191}
J.~Wu, Z.~Liu, Y.~Zhang, D.~Lei, B.~Zhang, and W.~Cao, ``Data-driven state of
  health estimation for lithium-ion battery based on voltage variation
  curves,'' \emph{Journal of Energy Storage}, vol.~73, p. 109191, 2023.

\bibitem{LAI2023128971}
X.~Lai, Y.~Yao, X.~Tang, Y.~Zheng, Y.~Zhou, Y.~Sun, and F.~Gao, ``Voltage
  profile reconstruction and state of health estimation for lithium-ion
  batteries under dynamic working conditions,'' \emph{Energy}, vol. 282, p.
  128971, 2023.

\bibitem{9858904}
S.~Bamati and H.~Chaoui, ``Developing an online data-driven state of health
  estimation of lithium-ion batteries under random sensor measurement
  unavailability,'' \emph{IEEE Transactions on Transportation Electrification},
  vol.~9, no.~1, pp. 1128--1141, 2023.

\bibitem{DRISCOLL2022104584}
L.~Driscoll, S.~de~la Torre, and J.~A. Gomez-Ruiz, ``Feature-based lithium-ion
  battery state of health estimation with artificial neural networks,''
  \emph{Journal of Energy Storage}, vol.~50, p. 104584.

\bibitem{YAO2023106437}
X.-Y. Yao, G.~Chen, M.~Pecht, and B.~Chen, ``A novel graph-based framework for
  state of health prediction of lithium-ion battery,'' \emph{Journal of Energy
  Storage}, vol.~58, p. 106437, 2023.

\bibitem{WEI2023108947}
Y.~Wei and D.~Wu, ``Prediction of state of health and remaining useful life of
  lithium-ion battery using graph convolutional network with dual attention
  mechanisms,'' \emph{Reliability Engineering \& System Safety}, vol. 230, p.
  108947, 2023.

\bibitem{TANG2023107734}
A.~Tang, Y.~Jiang, Q.~Yu, and Z.~Zhang, ``A hybrid neural network model with
  attention mechanism for state of health estimation of lithium-ion
  batteries,'' \emph{Journal of Energy Storage}, vol.~68, p. 107734, 2023.

\bibitem{BOCKRATH2023120307}
S.~Bockrath, V.~Lorentz, and M.~Pruckner, ``State of health estimation of
  lithium-ion batteries with a temporal convolutional neural network using
  partial load profiles,'' \emph{Applied Energy}, vol. 329, p. 120307, 2023.

\bibitem{KONG2021120114}
J.-z. Kong, F.~Yang, X.~Zhang, E.~Pan, Z.~Peng, and D.~Wang,
  ``Voltage-temperature health feature extraction to improve prognostics and
  health management of lithium-ion batteries,'' \emph{Energy}, vol. 223, p.
  120114.

\bibitem{LI2021116410}
Y.~Li, K.~Li, X.~Liu, Y.~Wang, and L.~Zhang, ``Lithium-ion battery capacity
  estimation --- a pruned convolutional neural network approach assisted with
  transfer learning,'' \emph{Applied Energy}, vol. 285, p. 116410, 2021.

\bibitem{LIN2022230774}
M.~Lin, D.~Wu, J.~Meng, J.~Wu, and H.~Wu, ``A multi-feature-based multi-model
  fusion method for state of health estimation of lithium-ion batteries,''
  \emph{Journal of Power Sources}, vol. 518, p. 230774, 2022.

\bibitem{WANG2023120808}
Z.~Wang, F.~Yang, Q.~Xu, Y.~Wang, H.~Yan, and M.~Xie, ``Capacity estimation of
  lithium-ion batteries based on data aggregation and feature fusion via graph
  neural network,'' \emph{Applied Energy}, vol. 336, p. 120808.

\bibitem{APICELLA2023109867}
A.~Apicella, F.~Isgrò, A.~Pollastro, and R.~Prevete, ``Adaptive filters in
  graph convolutional neural networks,'' \emph{Pattern Recognition}, vol. 144,
  p. 109867, 2023.

\bibitem{REN2023126590}
Y.~Ren, Z.~Li, L.~Xu, and J.~Yu, ``The data-based adaptive graph learning
  network for analysis and prediction of offshore wind speed,'' \emph{Energy},
  vol. 267, p. 126590, 2023.

\bibitem{7837898}
Y.~Zhu, Z.~Zimmerman, N.~S. Senobari, C.-C.~M. Yeh, G.~Funning, A.~Mueen,
  P.~Brisk, and E.~Keogh, ``Matrix profile ii: Exploiting a novel algorithm and
  gpus to break the one hundred million barrier for time series motifs and
  joins,'' in \emph{2016 IEEE 16th International Conference on Data Mining
  (ICDM)}, 2016, pp. 739--748.

\bibitem{Attia2020}
P.~M. Attia, A.~Grover, N.~Jin, K.~A. Severson, T.~M. Markov, Y.-H. Liao, M.~H.
  Chen, B.~Cheong, N.~Perkins, Z.~Yang \emph{et~al.}, ``Closed-loop
  optimization of fast-charging protocols for batteries with machine
  learning,'' \emph{Nature}, vol. 578, no. 7795, pp. 397--402.

\bibitem{10098615}
K.~Q. Zhou, Y.~Qin, and C.~Yuen, ``Lithium-ion battery state of health
  estimation by matrix profile empowered online knee onset identification,''
  \emph{IEEE Transactions on Transportation Electrification}, pp. 1--1., 2023.

\bibitem{SON2022121712}
S.~Son, S.~Jeong, E.~Kwak, J.-h. Kim, and K.-Y. Oh, ``Integrated framework for
  soh estimation of lithium-ion batteries using multiphysics features,''
  \emph{Energy}, vol. 238, p. 121712, 2022.

\bibitem{deng2021graph}
A.~Deng and B.~Hooi, ``Graph neural network-based anomaly detection in
  multivariate time series,'' in \emph{Proceedings of the AAAI conference on
  artificial intelligence}, vol.~35, pp. 4027--4035.

\bibitem{8215529}
C.-C.~M. Yeh, N.~Kavantzas, and E.~Keogh, ``Matrix profile vi: Meaningful
  multidimensional motif discovery,'' in \emph{2017 IEEE International
  Conference on Data Mining (ICDM)}, 2017, pp. 565--574.

\bibitem{10027730}
M.~Shahcheraghi, R.~Mercer, J.~M. De~Almeida~Rodrigues, A.~Der, H.~F.~S.
  Gamboa, Z.~Zimmerman, and E.~Keogh, ``Matrix profile xxvi: Mplots: Scaling
  time series similarity matrices to massive data,'' in \emph{2022 IEEE
  International Conference on Data Mining (ICDM)}, 2022, pp. 1179--1184.

\bibitem{8970797}
K.~Kamgar, S.~Gharghabi, and E.~Keogh, ``Matrix profile xv: Exploiting time
  series consensus motifs to find structure in time series sets,'' in
  \emph{2019 IEEE International Conference on Data Mining (ICDM)}, 2019, pp.
  1156--1161.

\bibitem{Yeh2019}
C.-C.~M. Yeh, Y.~Zhu, H.~A. Dau, A.~Darvishzadeh, M.~Noskov, and E.~Keogh,
  ``Online amnestic dtw to allow real-time golden batch monitoring,'' in
  \emph{25th ACM SIGKDD International Conference on Knowledge Discovery $\&$
  Data Mining}, July, pp. 2604--2612.

\bibitem{9589504}
K.~Q. Zhou, Y.~Qin, B.~P.~L. Lau, C.~Yuen, and S.~Adams, ``Lithium-ion battery
  state of health estimation based on cycle synchronization using dynamic time
  warping,'' in \emph{IECON 2021 -- 47th Annual Conference of the IEEE
  Industrial Electronics Society}, 2021, pp. 1--6.

\end{thebibliography}

\vspace{-11mm}

\vspace{12pt}
\end{document}